%% file: article.tex
\documentclass[journal]{IEEEtran}
\usepackage{amsmath,amsfonts,amssymb}
\usepackage{algorithmic}
\usepackage{algorithm}
\usepackage{array}
\usepackage[caption=false,font=normalsize,labelfont=sf,textfont=sf]{subfig}
\usepackage{textcomp}
\usepackage{stfloats}
\usepackage{url}
\usepackage{verbatim}
\usepackage{graphicx}
\usepackage{cite}
\usepackage{xspace}
\usepackage{color, soul}
\usepackage{booktabs}
\usepackage[absolute,showboxes]{textpos}
\input{macros.tex}

\begin{document}
\title{Combating the Effects of Cyber-Psychosis: \\Using Object Security to Facilitate Critical Thinking}
\author{
    \IEEEauthorblockN{Robert H. Thomson\textsuperscript{1}, Quan Nguyen\textsuperscript{2}, Essien Ayanam\textsuperscript{2}, Matthew Canham\textsuperscript{3}, Thomas C. Schmidt\textsuperscript{4}, Matthias W\"ahlisch\textsuperscript{5} \& Eric Osterweil\textsuperscript{2}} \\
    \IEEEauthorblockA{\textsuperscript{1}Army Cyber Institute, U.S. Military Academy, West Point, USA, \texttt{robert.thomson@westpoint.edu} \thanks{This paper was funded in part from an Office of Naval Research FY21 Multi-University Research Award (N0001423MP00219 \& N0001425MP00064) to RT.}} \\

    \IEEEauthorblockA{\textsuperscript{2}Department of Computer Science, George Mason University, Fairfax, USA, \texttt{\{qnguye31,eayanam,eoster\}@gmu.edu}
    \thanks{This work has been submitted to the IEEE for possible publication. Copyright may be transferred without notice, after which this version may no longer be accessible.}} \\ 
    \IEEEauthorblockA{\textsuperscript{3}Cognitive Security Institute, \texttt{matthew@cognitivesecurity.email}}\\    \IEEEauthorblockA{\textsuperscript{4}HAW Hamburg, Hamburg, Germany, \texttt{t.schmidt@haw-hamburg.de}} \\ 
    \IEEEauthorblockA{\textsuperscript{5}TU Dresden, Dresden, Germany, \texttt{m.waehlisch@tu-dresden.de}} \\ 
}

\maketitle

\begin{abstract}
  Humanity is currently facing an existential crisis about the nature of truth and reality driven by the availability of information online which overloads and overwhelms our cognitive capabilities, which we call Cyber-Psychosis. The results of this Cyber-Psychosis include the decline of critical thinking coupled with deceptive influences on the Internet which have become so prolific that they are challenging our ability to form a shared understanding of reality in either the digital or physical world. Fundamental to mending our fractured digital universe is establishing the ability to know where a \emph{digital object} (\ie a piece of information like text, audio, or video) came from, whether it was modified, what it is derived from, where it has been circulated, and what (if any) lifetime that information should have. %
  Furthermore, we argue that on-by-default object security for genuine objects will provide the necessary grounding to support critical thinking and rational online behavior, even with the ubiquity of deceptive content. To this end, we propose that the Internet needs an object security service layer.
  This proposition may not be as distant as it may first seem. Through an examination of several venerable (and new) protocols, we show how pieces of this problem have already been addressed. While interdisciplinary research will be key to properly crafting the architectural changes needed, here we propose an approach for how we can already use fallow protections to begin turning the tide of this emerging Cyber-Psychosis today!
\end{abstract}


\section{Introduction}
\IEEEPARstart{I}n the last decade, the average adult makes upwards of 35,000 decisions per day \cite{dubey2019towards}, up from an estimate of around 9,000 during working hours in 1994~\cite{wilson1994rop}.
Among key differences in the last 30 years are the \emph{sources} and fidelity of information and evidence.  This shift has been fueled by the rise of the Internet, search engines, social media, and online shopping. Exponentially more (lower quality) information is readily available to anyone with an Internet connection. While initially seen as a beneficial source for the democratization of information \cite{wihbey2014challenges}, this notion has been challenged by the rise of \textit{deceptive content} \cite{broniatowski2024role}, algorithmic feeding of news creating \textit{filter bubbles} \cite{kitchens2020understanding} causing echo chambers \cite{zimmer2019echo} and polarization \cite{thomson2024comparing}, and the paywalling of reliable academic content leading to the relatively greater availability of less rigorous research online \cite{nechushtai2019kind}. 

At the same time, there is evidence that increased use of the Internet and search engines - also known as the \textit{Google Effect}\cite{sparrow2011google} which is also being exacerbated by recent use of generative artificial intelligence (genAI) in education \cite{lee2025impact}, has led to a reduction in critical thinking and a failure of metacognitive calibration (e.g., an overconfidence in one's depth of understanding of a given topic; \cite{thomson2024investigating}). The sheer scale of information available means that it is possible to find a contrasting opinion on nearly any topic.
The information we consume almost exclusively takes form of \emph{``digital objects.''}
These include images, news stories, videos, audio recordings, and more.
They have their own lifecycles, origins, and targeting algorithms.
While many people believe that humans forage for information through a sensemaking process \cite{thomson2015general}, in the Information Age it is information which is algorithmically targeted to us \cite{kitchens2020understanding}. 
In a very real sense, data in digital objects should be reasoned about as if it has its own motivations \cite{bull2021dismantling}, or its own \emph{``agency.''}  
When one additionally factors in the opacity of the algorithms which feed us information, it becomes nearly impossible to judge the veracity of a claim. As such, we argue that the current information space leads to \emph{``Cyber-Psychosis,''} which will only become more problematic as we increasingly base our real-life decisions on information in digital objects and as technology pushes us into the metaverse \cite{wang2022survey}.

In this paper, we argue that there \emph{is} a viable and immediate path toward restoring people's ability to discern fact from fiction online. We illustrate that \emph{critical thinking} skills together with \emph{object security} can combat the roots of Cyber-Psychosis. Further, basic object security primitives have existed in the core of the Internet for decades, in protocols ranging from Secure/Multipurpose Internet Mail Extensions (S/MIME)~\cite{rfc8551}, to the Resource Public Key Infrastructure (RPKI)~\cite{rfc6480}, and the more recent Coalition for Content Provenance and Authenticity (C2PA)~\cite{c2pa}. Moreover, a combination of \emph{existing} general-by-design cybersecurity protocols and protections, such as the Domain Name System's (DNS')~\cite{mockapetris-sigcomm88} Security Extensions (DNSSEC)~\cite{rfc4033,rfc4034,rfc4035,yang2010deploying} and the DNS-based Authentication of Named Entities (DANE)~\cite{rfc6698,rfc7671,rfc8162,osterweil2020cybersecurity} are poised and ready to be used and built upon. As such, we issue a \textbf{call-to-arms} to develop interdisciplinary frameworks to help restore the security and resilience of our cognitive processes from deceptive online content. We argue that a first step includes evolving Internet services to produce secure-by-default digital objects via a generalized \emph{object security service layer}. This strategic goal will require basic research and engineering, but we also show that many of the most basic necessary components for us to address our tactical needs are already deployed and can make a meaningful difference if used today.

In the remainder of this paper, we begin by operationalizing the term Cyber-Psychosis and show how the present information space parallels the effects of clinical \textit{psychosis} in Section~\ref{sec:cybpsyc}. 
We then proceed, in Section~\ref{sec:critthinking}, to evaluate critical thinking frameworks used in education as a way to use reasoning strategies to ground out decision-making processes, and show a common set of principles which would mitigate the effect of Cyber-Psychosis. 
Using these critical thinking frameworks for guidance, we present a mapping between what they need, and how we can be (and in some cases already are) able to secure these with Internet object security in Section~\ref{sec:mapping}.
Next, in Section~\ref{sec:objs}, we propose a structured research agenda to bridge the gap between the existing protocol-specific object security service sets 
and research needed to realize a \emph{generalized} object security service layer for the Internet.
Then, we detail several prime example protocols which have already operationalized important aspects of generalized object security in Section~\ref{sec:past}.
Finally, we present a discussion of the state of our Internet in Section~\ref{sec:disc} and then conclude with our call-to-arms in Section~\ref{sec:conc}.

\section{Defining Cyber-Psychosis}\label{sec:cybpsyc}
In psychiatry, the hallmark of a psychotic episode is the loss of one's sense of reality \cite{ventura2010disorganization}. Cyber-Psychosis has been a term often used in science fiction to describe a mental condition that arises from excessive interaction with our dependence on technology, and often results in some kind of dehumanization. Originally popularized by William Gibson's \textit{Neuromancer} \cite{gibson2019neuromancer} and more recently by the video game CyberPunk 2077 \cite{steele2024chippin}, the concept involves people losing touch with reality due to their immersion in digital environments or cybernetic enhancements. While originally being an allegory for dehumanization and loss of autonomy by technology \cite{ross1991strange}, we argue that Cyber-Psychosis should actually be defined in line with the clinical definition for psychosis, which includes: 

\begin{enumerate}
    \item \textbf{Delusions:} Strongly held false beliefs that are resistant to reason or contrary evidence. In the extreme case, these can include paranoia and grandiose thoughts. 
    \item \textbf{Hallucinations:} Sensory experiences that appear real but are created by the mind. The most common type is auditory hallucinations, such as hearing voices that others do not hear.
    \item \textbf{Disorganized Thinking:} This can manifest as incoherent speech, difficulty organizing thoughts, or jumping from one topic to another without logical connection. 
    \item \textbf{Negative Symptoms:} These include diminished emotional expression, lack of motivation, and withdrawal from social activities.
\end{enumerate}

\noindent To best address these parallels, we will discuss each feature of psychosis in turn.

\subsection{Delusions} 
A common discussion in influence and media is the debate surrounding polarization in online spaces. Polarized thought reflects a strongly-held belief that is resistant to counter-arguments and contrary evidence. In fact, the contested \textit{backfire effect} \cite{nyhan2021backfire,wood2019elusive,swire2020searching} occurs when a person's belief in the polarized topic actually strengthens when presented with sound contradictory evidence. Furthermore, the \textit{continued influence effect} \cite{johnson1994sources,lewandowsky2012misinformation,ecker2021can} occurs when discredited prior beliefs continue to influence present thought, which has become prevalent in the current age of misinformation. These distortions in rational thought may be exacerbated by the Google effect insofar as people are overconfident in their understanding of a given topic, and there will almost always be some {digital object containing a} piece of confirming evidence on the Internet from which to anchor their prior beliefs.

\subsection{Hallucinations}
While the hallmark of hallucination is a sensory experience which appears real but isn't, we argue that a metaphorical hallucination occurs during the sensemaking process when evaluating information on the Internet. While it seems like it is the user who is searching for information, the underlying platform's algorithms are selectively feeding data back to the user. These algorithms are often opaque to the user, and this underlying agency of the algorithm {is applied to the digital objects whose data} may serve to steer the user to the algorithm's own aims (that of the company which made the algorithm) \cite{bull2021dismantling}. The focus of social media companies is to feed more ads to generate more revenue, thus the primary goal of information is not to satisfy the users request, but instead to maximize the user's continued interaction with the platform to generate more revenue. These algorithms effectively act as a hidden voice steering the user to a goal not of the user's own volition. This may only get worse with the advent of the metaverse and further integration of augmented and virtual reality into our daily lives.  
\subsection{Disorganized Thinking}
In the present discussion, disorganized thinking somewhat overlaps with our other analogies between clinic psychoses and the symptoms of Cyber-Psychosis. The hallmark of disorganized thinking is a difficulty in organizing thoughts and jumping between contexts without an obvious logical connection. For Cyber-Psychosis, this could be a result from the burnout and adrenal fatigue associated with Western society's \textit{always-on} work culture, leading to maladaptive behaviors such as doom-scrolling \cite{neijzen2024epistemic,sharma2022dark} and outrage fatigue \cite{crockett2017moral}. While not necessarily rising to the level of clinical significance in isolation, it exacerbates the other three features of Cyber-Psychosis discussed in this paper, and this disordered thought and failure of metacognition \cite{thomson2024investigating} makes it challenging for the user to detect this impairment.
 
\subsection{Negative Symptoms}
Perhaps the hallmark example of social and emotional withdrawal including a lack of motivation, comes from Hikikomori syndrome \cite{roy2021mental}. This syndrome, traditionally associated with Japanese culture but is seen around the world, reflects an extreme social isolation, emotional distress, lack of motivation, and addiction to digital spaces. 
This syndrome has not been officially classified as a mental disorder, but has been associated with prodromal psychosis, the sub-clinical phase predating a full psychotic illness \cite{stip2016internet}. A challenge with this syndrome is that there are limited effective evidence-based treatments \cite{sales2023hikikomori}. Most relevant to the current discussion is the role that digital spaces often have in the expression (and likely contributing factor to) this disorder. In the extreme case, the digital world becomes the person's reality.
In such cases, being able to quantitatively evaluate the veracity, origins, and provenance of the digital objects that shape a person's world becomes a critical necessity.

\subsection{Consequences of Cyber-Psychosis}
We argue that the Internet as currently designed and administered leads to the aforementioned four consequences as seen in numerous bodies of literature, which maps quite well onto the clinic definition for psychosis; thus we describe the current Internet age as leading to Cyber-Psychosis. This not only has a social component, there is evidence that extensive use of digital spaces and technology reduces the level of dopamine \cite{han2007dopamine} and interferes with the effectiveness of the limbic system \cite{huang2017autonomic}. This goes beyond pure Internet addiction as a compulsive behavior, it also impacts how we process reality itself. Given the availability of online information that goes well beyond our brain's ability to process \cite{schmitt2018too}, this \textit{information overload} can make it difficult to assess the veracity of information{, and we argue object security is a necessary first step}. There is evidence that Russia has weaponized the information overload and outrage fatigue facets of Cyber-Psychosis to interfere in foreign elections \cite{simchon2022troll}.

Given critical thinking frameworks were developed to assess the veracity of information, how well do they work {with digital objects on today's Internet} to mitigate the effects of Cyber-Psychosis? Users who have high literacy in algorithmic sources (e.g., chatbots) are more likely to trust their output and use them in their workflow \cite{shin2022people}, however this use may lead to over-trust \cite{schoenherr2024ai,Thomson2020KITT} and a reduction in critical thinking \cite{lee2025impact}. While some inoculation (pre-bunking) strategies have proven effective in limiting the acceptance of misinformation specifically \cite{kozyreva2020citizens}, we argue that this is insufficient to address the underlying issues of overload and fatigue which plague online spaces. 
We need object security to allow users to reason about information as \emph{it} encounters and is steered to them.

\section{Critical Thinking Frameworks}\label{sec:critthinking}
A review of recent frameworks endorsed by teachers and librarians (see Table \ref{tab:framework}) highlights a focus on critical reasoning, and it highlights three commonalities: 1) Review the source's authority, expertise, and motivations; 2) review the accuracy and soundness of the argument itself; and 3) understand the motivation behind why the argument exists. Perhaps the most well-studied is CRAAP, which was used extensively in the pre-Internet era to review primarily scientific claims. A concern with this methodology is that it focused primarily on the  article itself without reviewing other articles. More recent examples (e.g., CCOW) focus on situating the article in a broader world view, which mitigates some of these concerns and presents the opportunity to understand the context and motivations behind why this information was available to users in the first place \cite{liu2021moving}. 

These materials serve an important purpose by highlighting  factors for users to focus on when critically assessing a given argument. 
However, there are some concerns about these processes which may make them more counter-productive than would be seen at first glance. Namely, the focus on source authority and trustworthiness as an early indicator in most frameworks - over the logical form of the argument - may lead users to fall prey to source authority biases (e.g., appeals to authority, homophily-induced re-appraisals). For instance, Schoenherr and Thomson \cite{Schoenherr2021} (see also \cite{thomson2024investigating}) found that identifying the source of an argument as a scientist vs a layperson determined how effective participants were at evaluating the validity of an argument (i.e., its structure having a proper logical form).
This went beyond belief bias; the fact that we tend to judge the conclusion of an argument based on how well it fits with our prior beliefs instead of its logical structure. Even more troubling, the study found that participants' confidence scores were relatively unchanged when accuracy fell nearly 20\% when inducing belief bias, indicating that there was no dissonance or other metacognitive awareness that their critical thinking was impaired. Soprano et al \cite{soprano2024cognitive} further identified 39 cognitive biases that may influence fact-checkers in their critical thinking processes. Perhaps a better focus would be to relatively overweight techniques to evaluate the sources of the facts, not the source of the argument itself \cite{ajzen1991theory}?

\begin{table*}[ht!]
\centering
\begin{tabular}[t]{|p{2.3cm}|p{2.3cm}|p{2.3cm}|p{2.3cm}|p{2.3cm}|}
\hline
\textbf{CRAAP} & \textbf{CCOW} & \textbf{RADAR} & \textbf{ACLR} & \textbf{BEAM} \\
\hline
\textbf{Currency}: The timeliness of the information. & \textbf{Credibility}: The trustworthiness of the source. & \textbf{Relevance}: The importance of the information for your needs. & \textbf{Accuracy}: The correctness of the information. & \textbf{Background}: The context or history behind the source. \\
\hline
\textbf{Relevance}: The importance of the information for your research. & \textbf{Consistency}: Whether the information is consistent with other sources. & \textbf{Authority}: The qualifications or expertise of the source. & \textbf{Clarity}: The clearness and ease of understanding of the argument. & \textbf{Exhibit}: Evidence or examples presented by the source. \\
\hline
\textbf{Authority}: The qualifications and credibility of the author. & \textbf{Objectivity}: The lack of bias or impartiality in the source. & \textbf{Date}: The publication or last updated date. & \textbf{Logic}: The reasoning and structure of the argument. & \textbf{Argument}: The main claim or thesis the source is trying to prove. \\
\hline
\textbf{Accuracy}: The correctness of the content and the fact-checking process. & \textbf{Worldview}: Critical questioning of the source and its context as well as one's own views. & \textbf{Appearance}: The professional presentation and design of the source. & \textbf{Relevance}: The importance of the information in the given context. & \textbf{Method}: The approach or methodology used in the research. \\
\hline
\textbf{Purpose}: The reason the information exists and the intended audience. &  & \textbf{Reason}: The logic behind the conclusions or claims. &  &  \\
\hline
\end{tabular}
\caption{A list of common critical thinking frameworks. Others not shown here include DISCERN\cite{charnock1999discern}, Honcode \cite{boyer1998health}, and SIFT \cite{caulfield2021information}, among others.  Content derived from \cite{liu2021moving,tardiff2022have,sye2023tools,caulfield2021information}. It is a positive that CCOW highlights objectivity and worldview of the article and reader; as well as RADAR and ACLR highlighting the logical structure of the argument.}
\label{tab:framework}
\end{table*}

Another concern is that there is reasonable disagreement between methods.  Portillo et al \cite{portillo2021quality} found disagreement between critical thinking frameworks' (DISCERN, CRAAP, and HONcode) evaluation of common medical websites including Medline, Healthline, Mayo Clinic, and WebMD. Specifically Medline Plus was the only website which did not receive HONcode certification, while having the highest reliability by DISCERN and CRAAP. 

\subsection{Assessable Features of an Argument to Support Critical Thinking}

\begin{itemize}
    \item Source reputation of publication or website. Source reputation of the author.
    \item Data referenced is accurate. Are articles pointing to source material? Is evidence from reputable source?
    \item Logical consistency (valid and sound points being made). 
    \item Bias/Objectivity. Why is what being said being said. Who else is saying this. Who is saying the counter? Can there be a dashboard of opinions on a topic?
    \item Clarity and Precision. Is there language used which could (intentionally) lead to obscurity/vagueness
    \item Relevance: are examples relevant or are they meant to be evokative and to mislead towards affective red herrings and unrelated bits.
    \item Does it accept and refute a counter-claim (debating 101).
\end{itemize}

By evaluating the commonalities in these frameworks, we propose that cybersecurity protections can be applied to digital objects in order to allow Internet users to best understand the common features of an argument which support rational human decision-making and which may help mitigate the effects of Cyber-Psychosis.

\input{mapping.tex}
\input{objs.tex}
\input{past.tex}
\input{disc.tex}
\input{conc.tex}

\section*{Acknowledgment}
The views expressed in this work are those of the authors in their unofficial capacity and do not necessarily reflect the official policy or position of the U.S. Military Academy, U.S. Army, Office of Naval Research, or U.S. Government.

\bibliographystyle{IEEEtran}
\bibliography{references,rfc,paper}

\newpage

\begin{IEEEbiographynophoto}{Robert Thomson} is an Associate Professor in the Engineering Psychology Program at the United States Military Academy, a Principal Research Scientist at the Army Cyber Institute, and currently serves as the Dean's Fellow for Research. Dr. Thomson has over 13 years of post-graduate research experience and over 75 invited and refereed academic publications in the domains of computational modeling, intelligence analysis, cybersecurity, and artificial intelligence. He has been selected and supported more than \$25M in reimbursable research from IARPA, DARPA, NIH, the Office of Naval Research, and several Army organizations.
\end{IEEEbiographynophoto}
\begin{IEEEbiographynophoto}{Quan Nguyen} is a second-year Ph.D. student in Computer Science at the George Mason University (GMU) under the advising of Prof. Eric Osterweil (GMU). He is interested in topics in Cryptography and Internet Security. His research focuses on evaluating the longitudinal security of current Internet's object security protocols, e.g. DNSSEC, etc., and analyzing the lessons from the deployments of those protocols that can guide the design and implementation of the future Internet's object security protocols.
\end{IEEEbiographynophoto}
\begin{IEEEbiographynophoto}{Essien Ayanam} is a Ph.D. student in Computer Science at George Mason University (GMU). He has over 15 years of experience in designing and deploying secure network architectures, security policies, and procedures. He serves as a Senior Network Security Engineer in the Investigative Technology Division at the Drug Enforcement Administration, addressing critical security challenges in the government space. His research focus is on measurable security for Internet-based protocols, digital object protections, and automotive network technology (CAN). As a researcher at GMU’s Measurable Security Lab, he examined the impact of Internet protocols on security mechanisms and assisted with the development of a framework for evaluating digital object protections.

\end{IEEEbiographynophoto}
\begin{IEEEbiographynophoto}{Matthew Canham} is a former Supervisory Special Agent with the Federal Bureau of Investigation (FBI), he has a combined twenty-one years of experience in conducting human-technology and security research. He currently holds an affiliated faculty appointment with George Mason University, where his research focuses on threats posed by maliciously produced AI generated content and synthetic media social engineering. Dr. Canham recently founded the Cognitive Security Institute, a non-profit organization dedicated to understanding the key components of cognitive attacks and discovering the best ways to defend against these.
\end{IEEEbiographynophoto}
\begin{IEEEbiographynophoto}{Thomas C. Schmidt} is  a Professor of computer networks and Internet technologies with the Hamburg University of Applied Sciences, Hamburg, Germany, where he heads the Internet Technologies Research Group. He was the Principal Investigator in a number of EU, nationally funded, and industrial projects, as well as a Visiting Professor with the University of Reading, Reading, U.K. He is a cofounder and a coordinator of several successful open source communities. His current research interests include development, measurement, and analysis of large-scale distributed systems, such as the Internet or its offsprings.
\end{IEEEbiographynophoto}

\begin{IEEEbiographynophoto}{Matthias W\"ahlisch}  is a Full Professor and holds the Chair
of Distributed and Networked Systems at the Faculty of Computer Science at TU Dresden. He is also a Research Fellow of the Barkhausen Institut. His
research and teaching focus on scalable, reliable, and
secure Internet communication. This includes the design and evaluation of networking protocols and architectures, as well as Internet measurements and analysis. Matthias is involved in the IETF since 2005 and
co-founded multiple successful open source projects such as RIOT and the RTRlib.

\end{IEEEbiographynophoto}

\begin{IEEEbiographynophoto}{Eric Osterweil} is the Associate Director of the Center for Assurance Research and Engineering (CARE) and an Assistant Professor in the Department of Computer Science at George Mason University. He is the former co-Chair of the 2nd Security, Stability, and Resiliency (SSR2) Review Team in the ICANN Community. His research focuses on Internet-scale object security, cybersecurity protocols, and large-scale Internet measurements.  He has authored and co-authored over 130 papers, articles, patents, and RFCs and maintains the most complete longitudinal measurement corpus of the global deployment of DNSSEC, whose data spans from the protocol's final standardization in 2005 through to today.
\end{IEEEbiographynophoto}

\vfill

\end{document}

%% file: macros.tex
\usepackage[dvipsnames]{xcolor}

\usepackage{pifont}
\newcommand{\cmark}{\ding{51}}%
\newcommand{\xmark}{\ding{56}}%

\usepackage{ifthen}

\definecolor{verylightgray}{gray}{0.8}
\definecolor{__red}{rgb}{0.8,0.1,0.1}
\definecolor{__green}{rgb}{0.1,0.6,0.1}

\newcommand{\etc}{etc.\@\xspace}

\newcommand{\eg}{\textit{e.g.,}\@\xspace}
\newcommand{\ie}{\textit{i.e.,}\@\xspace}

\newcommand\para[1]{\vspace{0.05in} \noindent \textbf{#1}}

\newcommand{\one}{{(\bf I)}\xspace}
\newcommand{\two}{{(\bf II)}\xspace}
\newcommand{\three}{{(\bf III)}\xspace}
\newcommand{\four}{{(\bf IV)}\xspace}
\newcommand{\five}{{(\bf V)}\xspace}
\newcommand{\six}{{(\bf VI)}\xspace}

\newcolumntype{G}{>{\centering\let\newline\\\arraybackslash\hspace{0pt}}m{0.12\linewidth} }
\newcolumntype{L}{>{\centering\arraybackslash} m{0.3\linewidth} }
\newcolumntype{M}{>{\centering\arraybackslash} m{0.15\linewidth} }
\newcolumntype{S}{>{\centering\arraybackslash} m{0.08\linewidth} }

%% file: mapping.tex
\section{Combating Cyber-Psychosis with Object Security}\label{sec:mapping}

While modern Internet usage has shifted almost exclusively to \emph{digital objects},
these objects present a foundational challenge.
This is because existing cybersecurity protocols are most commonly only designed to protect data for the short windows of time while it is being transmitted, \ie \emph{``transmission security.''}
The most common example of this is the Transport Layer Security (TLS)~\cite{rfc8446} protocol.
However, one central security challenge posed by digital objects is that they exist beyond just when they are being transmitted.
They exist on various servers and Internet services, are shared between users, and are promoted by platforms' targeting algorithms.
Therefore, they must \emph{also} be secured while ``at rest.''
As a result, digital objects require a different type of security, which can provide more protracted protections, which we call \emph{``object security.''}
This frames a basic challenge of what protections we need for digital objects so as to best protect them in ways that will let users ingest them for critical thinking.

To know whether our objects are (or can be) secure, we must first understand precisely what protections are needed.
Using the critical thinking frameworks from Section~\ref{sec:critthinking} as representative examples, inspection of their objective requirements reveals several recurring objective needs.
Of the example set described in Table~\ref{tab:framework}, at least three of the five examples require \emph{origin authentication} protections (\ie being able to securely verify the sources of information).
Specifically, origin authentication would be necessary for CRAAP to establish ``Authority.''
Similarly, CCOW would need it for ``Credibility,'' ``Objectivity,'' and ``World View.''
RADAR would also need it for ``Authority.''
In addition, to implement these frameworks on today's Internet, four frameworks would also need combinations of integrity protections and \emph{process provenance}~\cite{10.1145/1453101.1453147}.
Lastly, at least two rely on forms of data \emph{lifecycle} management.
As origin authentication, provenance, and lifecycle emerge as common requirements, protocol-specific examples can be seen to exist in several prominent Internet protocols (detailed in Section~\ref{sec:past}).
However, what is \emph{also} needed is a basic understanding of what a generalized object security service layer for the Internet should look like.
That future definition looms large as an open research challenge.

%% file: objs.tex
\section{What Does it Mean to ``Secure'' Objects?}\label{sec:objs}

Understanding what is needed for generalized object security between separate administrative domains (\eg competing companies, collaborating universities, different nations, \etc) must be decomposed into basic pieces. 
For this, we propose a set of research questions to start:

\para{Research question~\one, 
what are the basic nature and requirements of objects' \emph{origin authentication} protections?}
This challenge will involve locating origins online and being able to securely verify that genuine objects have come from their reported sources.
To illustrate, consider an online new story that is shared multiple times, over multiple social media platforms.
Recipients should want to securely verify who authored it, \ie its \emph{origin}.
Further, that origin may not be useful for the recipient unless it is a name that she/he recognizes.
To accomplish this with public-key cryptography, object recipients will need to securely learn the public keys of object originators.
However, issues ranging from private key compromise to standard cryptographic hygiene dictate that keys will inevitably (if not periodically) need to be changed over time,
\ie have a \emph{lifecycle}.
When an origin needs to change its key it should not necessarily need to reissue all of its objects that are verified by that key.
As a result of this, the cryptographic keys, themselves, will not be sufficient as identifiers.
Identifying and locating inter-administrative data origins and mapping them to changing identifiers on the Internet has frequently been accomplished through unambiguous \emph{naming}.
In the context of origin authentication, these names must be semantically able to unambiguously locate public keys for identified object sources.
Research challenges include understanding what the necessary and sufficient natures of a usable namespace for digital objects.

\para{Research question~\two, 
what semantics will the \emph{lifetimes} of objects' protections need?}
Developing protection-focused guidance for \emph{object lifetime} will involve creation of evaluation frameworks to inform decisions about how long objects should be protected.
For example, some digital objects' content may only be intended to be valid for certain predictable periods of time.  
Therefore, should objects \emph{only} be verifiable in specific time frames?
By contrast, some objects' content may be intended to be valid indefinitely.
Should those objects have open ended verifiability?
If so, should their validities be periodically renewed, or semantically specified as indefinite?
During validity periods of objects' lifetimes, their data must have enforceable integrity protections.

\para{The above research question leads directly into question~\three, what does an effective relationship between objects' lifetimes and the \emph{lifecycles of the cryptographic keys} that protect and verify them look like?}
While an object needs to be validated, it must have a corresponding cryptographic key to do that verification and integrity protection.
This means keys cannot expire during periods of time that the objects they cover are expected to be verified by them.
Instead, keys must exist and be valid for the periods that objects need them, which forms an implicit requirement that their lifecycles be congruent.
Among the immediate research challenges this poses is how to manage situations in which a source's key needs to change (\eg emergency, planned retirement, \etc)? 
In situations like these, having objects' keys be located by names will facilitate key transitions~\cite{osterweil2022beginning}, but that alone will not guarantee continuity of protections.

\para{Research question~\four, 
what aspects of \emph{integrity} and \emph{provenance} will be needed?}
``Provenance'' is a word whose common usage often includes application to the concept of ``chain-of-custody'' (\eg what is the ownership history of a piece of art).
The word has also been a topic of a great deal of the computer science literature.
In the past, it has been used to document audit trails for e-science~\cite{simmhan2005survey,10.1145/1453101.1453147,sahoo2009provenir,miles2007requirements,zhao2004using}, and more recently a growing body of research has considered its applicability to combat deepfake images~\cite{rosenthol2022c2pa,10167902,sedlmeir2023battling,10.1145/3625468.3652198}.
In the more recent focus, much attention has been paid to documenting when an object has been altered (such as when images are touched-up or filtered).
In these cases, it must be possible to securely inspect and evaluate these changes.
We, therefore, cast these protections as a superset of \emph{integrity protections}, which are inherently offered by this type of object provenance.
However, while this type of provenance is necessary, it is not sufficient for combating \emph{misinformation and disinformation}.
We contend that it is also necessary to be able to verifiably express the ``process(es)'' by which objects have been \emph{derived}, or \emph{process provenance}~\cite{10.1145/1453101.1453147}.
Said differently, if an object represents a conclusion that is derived from other objects, it is necessary to securely reference these other objects directly.
This will likely entail the same origin naming requirement for each component of an object's provenance.
Further, with digital objects having their own agency on the Internet, knowing \emph{where} they have appeared can be as important as what they say, or where they originated (\ie chain of custody provenance).

\para{Research question~\five: 
will \emph{confidentiality} protections need different approaches?}
Whereas transmission security protocols have extensively protected confidentiality, are their models applicable to generalized object security?
As we discuss in Section~\ref{sec:past}, there are a number of existing protocols that have attempted aspects of an overall object security service layer.
While these examples illustrate applicable lessons for our work here, vanishingly few have attempted longitudinal confidentiality protections between separate administrative entities.
Does this imply that there are non-obvious complexities in designing confidentiality protections for inter-administrative object security, or perhaps that they are not necessary?

\para{Research question~\six, \emph{evaluability}, 
how can object creators and recipients \emph{know} if protections are remaining effective, or not?}
There needs to be a way for content authors, object recipients, and anyone else to \emph{know} if object protections have failed and/or if they have worked.
We believe that object security, in particular, will require a robust approach to knowing the protection status of objects.
This is, in part, because of the protracted periods of their protections.
For example, knowing that an object has valid cryptographic signatures at one point in time does not necessarily tell a relying party that it was protected yesterday, or will be tomorrow.
Moreover, if an object needs to remain secure for any critical function, being able to prove that it has or has not been secure will be necessary to deriving trust from~it.

This set of research tasks is an example of how the complex nature of object security could be decomposed.
Other valid decompositions could lead to different research tasks, but we believe the above set is an important example starting point.
However, there exist examples of inter-administrative security protocols on the Internet that have  operationalized aspects of these, which both illustrate existence proofs that these protections are attainable, and in some cases serve as \emph{usable} options for generalized object security \emph{today}.
We examine these in detail next, in Section~\ref{sec:past}.

%% file: past.tex
\section{Where have digital objects been secured?}\label{sec:past}

Some notable security protocols are designed to protect objects, such as
the Coalition for Content Provenance and Authenticity (C2PA)~\cite{c2pa}, the Secure/Multipurpose Internet Mail Extensions (S/MIME)~\cite{rfc8551}, Pretty Good Privacy (PGP)~\cite{zimmermanofficial}, the Domain Name System's (DNS')~\cite{mockapetris-sigcomm88} Security Extensions (DNSSEC)~\cite{osterweil2022beginning, rfc4033,rfc4034,rfc4035}, 
the Resource Public Key Infrastructure (RPKI)~\cite{rfc6480},
and Windows' Authenticode code signing protocol~\cite{authenticode}.
These examples range from a couple of years old to decades and their deployments offer insights into generalizable object security.
In this Section, we use our candidate research areas \one to \six as guideposts, and
Table~\ref{tab:synth} summarizes discussions of them.

{\small
\begin{table*}
  \caption{Implementation of object security protections that are available (\cmark), partially available ((\cmark)), or not available (\xmark)}\label{tab:synth}
  \centering
  \begin{tabular}{   l   c   c   c   c   c   c   c   }
    \toprule
    {\bf Protocol} 
    & {\bf Origin Authentication} 
    & {\bf Lifetime} 
    & {\bf Keys-to-Objects} 
    & {\bf Integrity} 
    & {\bf Provenance} 
    & {\bf Confidentiality} 
    & {\bf Evaluation} \\ 
    \midrule
    C2PA 
      & \xmark 
      & \xmark 
      & \xmark 
      & \cmark
      & (\cmark)
      & \xmark 
      & \xmark\\ %
    S/MIME \& PGP  
      & \xmark 
      & \xmark 
      & \xmark 
      & \cmark
      & \xmark
      & \cmark 
      & \xmark \\ %
    DNSSEC 
      & \cmark 
      & \cmark 
      & \xmark 
      & \cmark
      & \xmark
      & \xmark 
      & \xmark \\ %
    RPKI 
      & \cmark 
      & \cmark 
      & \xmark 
      & \cmark
      & \xmark
      & \xmark 
      & \xmark \\ %
    Authenticode  
      & \xmark 
      & \cmark 
      & \cmark 
      & \cmark
      & \xmark
      & \xmark 
      & \xmark \\ %
    \bottomrule
  \end{tabular}
\end{table*}
}

\begin{figure}
  \centering
  \includegraphics[width=0.45\textwidth]{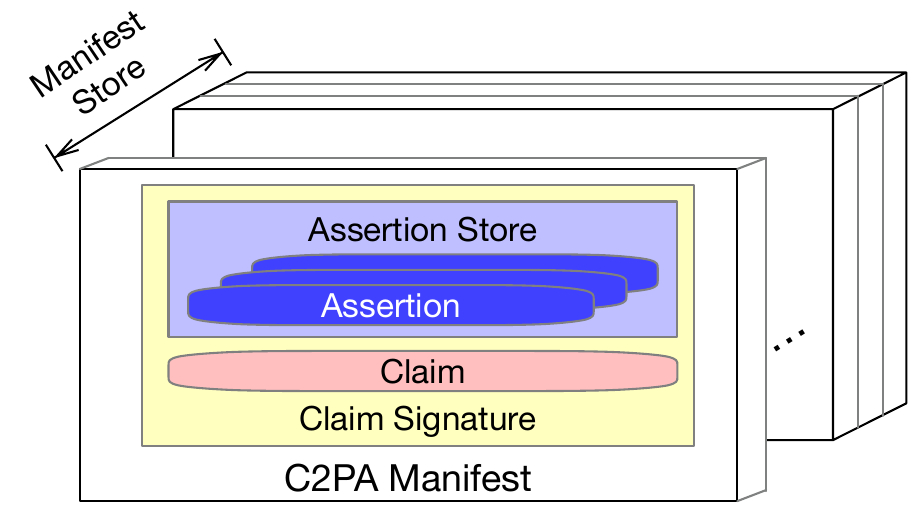}
  \caption{C2PA provenance and integrity is implemented by Manifests.}
  \label{fig:c2pa}
\end{figure}

\para{\bf C2PA:}
In recent years the C2PA has emerged as a leading organization in the provenance space~\cite{rosenthol2022c2pa,10167902,sedlmeir2023battling,10.1145/3625468.3652198}, \ie protecting objects like JPEG, PNG, and MPEG.
Its protections are focused on describing how media objects have been created and/or modified, and is implemented by a set of constructs called ``manifests'' (illustrated in Figure~\ref{fig:c2pa}).
Each manifest contains three to four data items: assertions, claims, a signature, and optionally W3C Verifiable Credentials (VCs)~\cite{w3c-vc}.
Assertions describe when an image object was taken, what camera took it, what editing tool may have modified it, and generally what action may have been taken on that object.
Claims bundle assertions together by a ``claim generator'' via cryptographic hashes.
They indicate who generated the claim, what software was used to generate it, \etc
Finally, the claim signature is generated over the whole manifest to vouch for its integrity.
Every modification to a media object should result in a new manifest.
The total set is accumulated into the object's ``manifest store,'' (from the ``origin manifest'' to the ``active manifest'').
The C2PA specification states that trust in objects should be determined by the identity of the signer.

The C2PA specification \emph{partially} implements origin authentication (our research area \one) through its use of X.509 certificates.
Verification of objects is accomplished by verifying claim signatures using their corresponding X.509 certificate.
However, verification \emph{of that certificate} hinges on its Subject Alternative Name (SAN), which is most commonly a DNS domain name.
This inherently forms a basic dependency on the DNS \emph{namespace}, but these names are not currently used to authenticate origins' keys.
C2PA media objects do not offer semantics for lifetime protections (our research area \two).
Any lifetime restrictions are (at best) implicit from certificate validity lifetimes.
This specification does not have an explicit relationship defined between media objects and the keys that protect them.
Thus, the key to object management (our research area \three) is not protected by C2PA.
Provenance is 
implemented using X.509 certificates, Concise Binary Object Representation (CBOR), and constructs called (JPEG Universal Metadata Box Formats) JUMBFs.
However, this is only one component of what we consider the \emph{process provenance} in our vision for research area \four.
The provenance of C2PA media objects describes their \emph{modification history} provenance, and includes integrity protections built in.
We propose that this type of modification history of digital objects 
is necessary, but not sufficient for problems like misinformation and disinformation.
This is why we propose the need for \emph{process provenance}, \ie which \emph{external} objects may have been involved (or led to) the creation of objects.
While images may not actually be the beneficiaries of this type of protection, they can be included in the process provenance for others (such as news stories derived from them).
The C2PA specification does not provide confidentiality (our research area \five), however recent research from the literature proposes an approach to provide this~\cite{10167902}.
Lastly, C2PA's protections do not include an evaluability approach (our research area \six).

\para{\bf S/MIME \& PGP:}
These protocols are among the oldest object focused inter-administrative security protections in use today, though their deployment traction has lagged.
They are very similar in the protections they offer.
Among the primary benefits derived from using these protocols today is their confidentiality protections (our research area \five).
These protocols both protect general object types, anything that can be sent in an email message.
However, they do not address the rest of our research areas.
Neither protocol implements origin authentication (our research area \one).
These protocols are not able to verify that keys are securely associated with the names they report.
They each rely on external mechanisms (X.509 for S/MIME and the Web of Trust for PGP), but neither serves as a reliable inter-administrative origin authentication solution, which we describe in detail in previous work~\cite{osterweil2020cybersecurity}.
Neither of them implement lifetime protections (area \two).
Neither have semantics or mechanisms to manage the lifecycles of keys to their protected mail objects (research item \three).
They both have integrity protections, but do not have provenance mechanisms (area \four).
Lastly, neither have an evaluability approach (area \six).

\begin{figure}
  \centering
  \includegraphics[width=0.45\textwidth]{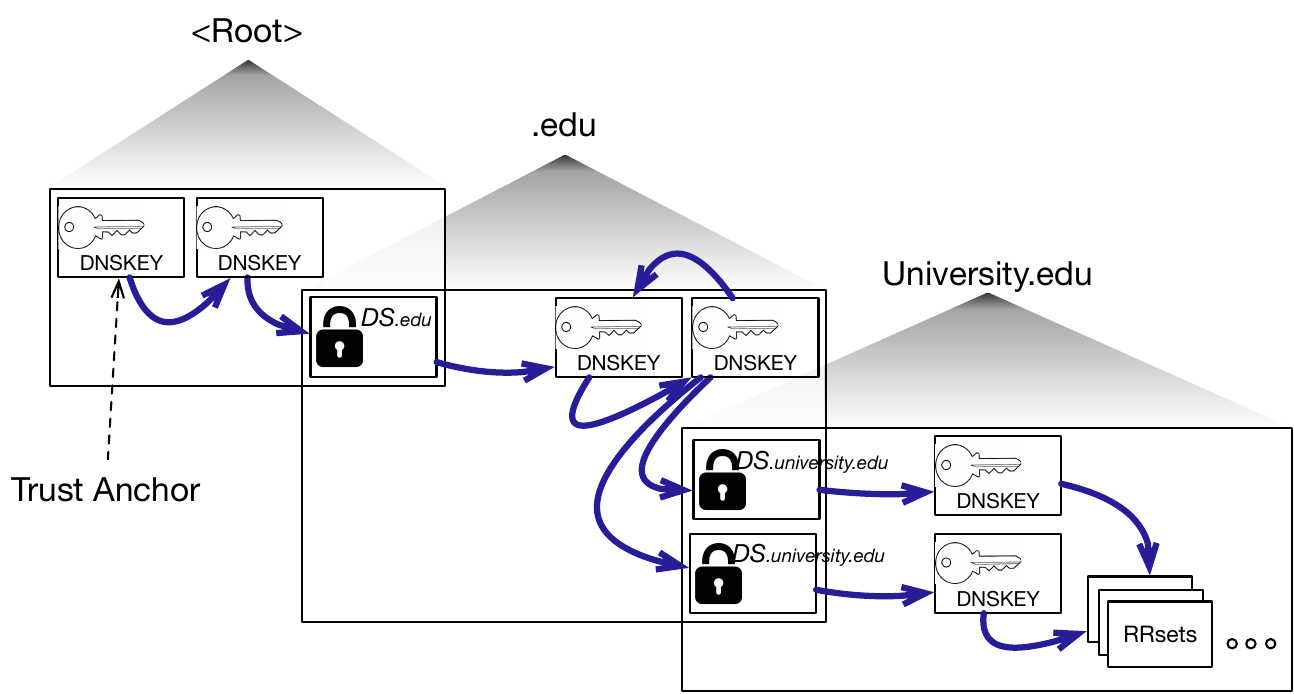}
  \caption{DNSSEC's hierarchical namespace enables secure key learning.}
  \label{fig:dnssec}
\end{figure}

\para{\bf DNSSEC:}
DNSSEC's namespace is general (the DNS is the Internet's de facto name mapping system), but its object type is DNS-specific (as opposed to being generally useful for other applications, like S/MIME and PGP).
In DNSSEC, zone names map to public keys ({\tt DNSKEY} records), and these keys are used to verify signatures attached to DNSSEC's canonical object type: the ``RRset,'' illustrated in Figure~\ref{fig:dnssec}.
Zone administrators generate signatures (in {\tt RRSIG} records) that  cover objects.
Research has shown that its deployment size has been growing steadily since its standardization~\cite{osterweil2022beginning, secspider-web}.
Measurements also show that the size of DNSSEC validator traffic (\ie clients) now exceeds 33\% of global DNS~\cite{apnic-validator-stats}.
Research has also demonstrated the robustness of the policy framework around the root zone and its key management~\cite{10.1145/3694809.3700746,Muller:2019:RRR:3355369.3355570}.

DNSSEC implements origin authentication (area \one).
Zones deploy their own named keys, zones also aggregate named objects.
These names exist in DNS' collision-free domain namespace.
In DNSSEC, the namespace itself binds keys to objects.
Though the DNS has been ubiquitous for decades, the ongoing growth and evolution of its ecosystem has exposed security challenges that relate to name collisions.
Past research~\cite{chen2016mitm} has illustrated that the introduction of new generic Top-Level Domains (gTLD) enabled Man-in-the-Middle (MitM) attacks because of name collisions.
Subsequent research~\cite{chen2017client} illustrated remediative actions for this that an object security service layer will need to incorporate.
DNSSEC's trials and tribulations over the complexity of securing a namespace should be advisory for alternate namespace proposals.
DNSSEC implements lifetime protections (area \two) through its signatures over objects.
Each signature specifies a definitive inception and expiration time.
The implication of this approach is that if any given object needs to be verifiable after its previously specified expiration, it will need to be re-signed with a new inception/expiration.
Conversely, if an object becomes invalid while it is still within its specific lifetime, there is no recourse for a zone administrator.
The object can be removed from an operational zone, but will still be replayable to validating resolvers and caches.
There exists no research literature studying the impact of object lifetimes in DNSSEC.
It does implement integrity protections over its objects (through its covering {\tt RRSIG}s), but it does not implement provenance protections (area \three).
Research area \four, lifecycle alignment between keys and objects, has begun to be be studied in the literature.
In~\cite{osterweil2022beginning} we found that there is a large disconnect between the prescribed processes for key management and the actual operationally deployed processes.
One broader lesson is that protocols need to be designed to maintain continuity of \emph{protections} for covered objects.
Research area \five (confidentiality) is not implemented by DNSSEC.
Though, there has been a push for protocols called ``encrypted DNS,'' such as DNS over HTTPS (DoH), DNS over TLS (DoT), \etc
These, however, implement transmission security and do not address object security protections.
Research addressing evaluation of DNSSEC (area \six) has primarily focused on its availability, rather than its object protections.
For example, a challenge that was identified by research was a reduced incidence of validation by resolvers (clients)~\cite{chung2017longitudinal,182950}.
This research discussed the success rates of validation, but did not examine the success rates of validating the \emph{signed objects}.
This is not a critique of the research and does not reflect a limitation of it.
Rather, this is an observation that this research's lessons are not focused on the \emph{object security} implications, but instead the overall deployment's performance.

\para{\bf RPKI:}
This protocol secures the global routable Internet Protocol (IP) address space and the Autonomous System Number (ASN) space.
To compare this with the DNS, names are IP prefixes and the canonical object type is called a Route Origin Attestation (ROA), each of which embeds an End Entity (EE) certificate.
The ``origin'' in ROAs refers to a BGP origin ASN, not the creator of an object.
RPKI defines a BGP-level verification process called Route Origin Validation (ROV)~\cite{RFC6811} to verify the correctness of routing information in ROAs.
The RPKI has been undergoing deployment for over 10 years~\cite{rodday2023resource}, and recent 
research shows its adoption and deployment traction have been growing~\cite{10.1145/3618257.3624806} since the beginning \cite{wahlisch2015ripki,10.1145/3211852.3211856}.

The RPKI implements origin authentication (area \one) by creating a hierarchy of digital certificates that chain from a Regional Internet Registry (RIR) to any ROA.
This means that any rightful resource holder of an IP prefix can create a ROA to provably attest to a mapping of that prefix to an ASN.
Complementing this, any relying party can verify a ROA by bootstrapping all RIR trust anchors and recursively following prefix delegation certificates to a ROA's EE certificate.
The lifetime protections (area \two) of RPKI are enforced by definitive lifetimes of ROAs that are inherited from their certificate delegation chains (\ie certificate lifetimes).
Similar, but distinct, is that ROAs can indicate their desired cache control timings.
This is, however, distinct from the object protections.
Studies on lifetimes of ROAs (\ie object security studies) are absent from the literature.
Similarly, research area \three---alignment of key and object lifecycles---is absent from the protocol and the literature.
While the EE certificates contained in ROAs could (potentially) be aligned with the intended ROA lifetime, the \emph{overall} verifiable lifetime of the ROA depends on the \emph{full delegation chain} (which is decoupled from ROA lifetimes).
The RPKI implements integrity protections over its objects, but does not provide a provenance mechanism (area \four).
This, likely, is not an omission but a derivative of the fact that ROAs should not be derived from a source other than the rightful holder of a resource.
Similarly, the RPKI does not have a motivation for the confidentiality of its objects (area \five).
Finally
however, as with DNSSEC, the security of the \emph{objects themselves} remains un-evaluated (area \six).
Calculating ROV status, though important, is not equivalent to evaluating if the objects (ROAs) themselves are being protected by the RPKI system (ROV $\ne$ ROA protections).
ROV evaluates the correctness of the routing announcements, but it is not designed evaluate if the object itself is, or has remained, secure.
For example, it does not have semantics to protect against newer objects that supersede older objects by the same ``name,'' or against objects that have been otherwise invalidated.
As with DNSSEC, this is not a critique of the research and does not reflect a limitation of it.

\begin{figure}
  \centering
  \includegraphics[width=0.45\textwidth]{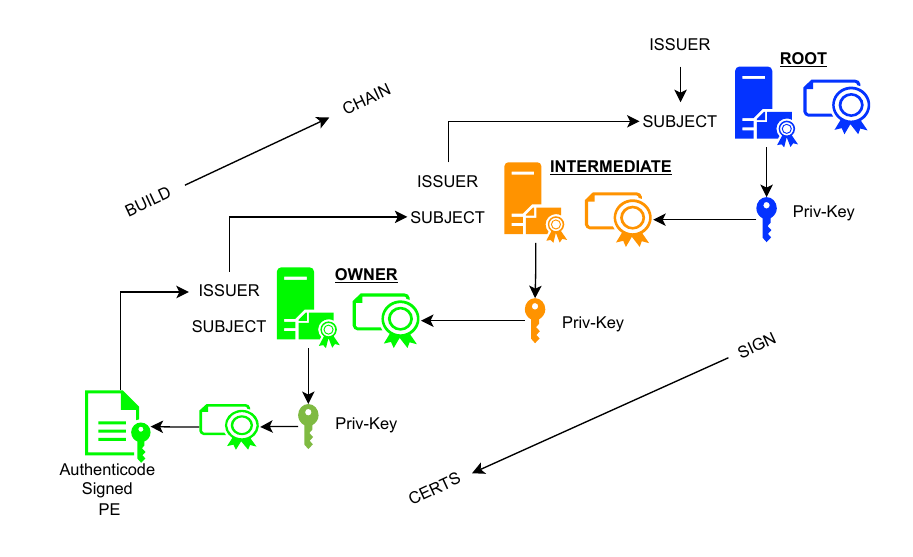}
  \caption{Authenticode protects objects through an X.509 delegation hierarchy.}
  \label{fig:authenticode_v2}
\end{figure}

\para{\bf Authenticode:}
This protocol is designed to prove the authenticity of code objects, \ie Portable Executable (PE) files.
This includes executables (.exe), dynamically loaded libraries (.dll), and drivers (.sys). 
It is primarily designed to verify the origin and integrity of software binaries, as depicted in Figure~\ref{fig:authenticode_v2}. 

Authenticode attempts to implement origin authentication (area \one) via an embedded digital signature, which ties each PE object to its ``owner’s certificate.''
However, without a formal name lookup for the PE file itself or verification of the certificate's name, it does not fully validate the origin's identity.
It only verifies the binding between a PE object and the cryptographic key in the certificate.
Another key aspect of Authenticode is its signature lifecycle, which connects to research area \two (lifetime protections). 
Authenticode optionally supports timestamping, extending signatures' validity beyond the expiration of the signing certificates. 
When implemented, timestamping grants the signing certificate and signature indefinite validity. 
This approach aligns the lifecycle of the digital signature with the signing certificate. 
While timestamping addresses the issue of expiring certificates and signatures, it introduces a potential vulnerability: indefinitely valid certificates may remain exploitable if compromised.
Authenticode implements a technique to align object lifetimes with key lifecycles (area \three) by explicitly decoupling them, by default.
When a PE is created, software can obtain a signed attestation from a Time Stamp Authority (TSA) that attests that all cryptographic material was valid at creation time.
Then, all certificates in the delegation chain are bundled into the PE (with their definitive expiration times included).
This approach allows verifying software to use expired certificates iff they were valid at creation time, thereby ``securely'' decoupling their lifecycles from object lifetimes.
Authenticode emphasizes integrity but makes no claims regarding provenance (research area \four).
it ensures integrity by hashing the signed PE file. 
Any modification to the file after signing can be detected by comparing the computed hash during verification. 
Authenticode does not address the confidentiality of PE files (area \five), focusing exclusively on owner authenticity and integrity. 
Similarly, the security of the PE files themselves (area \six) is unaddressed by the specification.

%% file: disc.tex
\section{Discussion of the State of the Internet}\label{sec:disc}

In the pursuit of a generalized object security service layer, Table~\ref{tab:synth} illustrates that several protocol-specific incarnations of necessary security services and protections have already been successfully deployed at scale in the Internet.
However, what is also deployed are sets of general-by-design object protections, lying fallow.

\para{\bf Object Security Services:}
Across the object security protocols we surveyed, roughly half operated on objects that were protocol-specific (\eg RPKI, DNSSEC, and Authenticode). 
However, S/MIME, PGP, and C2PA are designed for protections over generic object types.
Similarly, DNSSEC implements a general namespace (the DNS') for the Internet's entire array of online services.

For origin authentication, we argue that \emph{successful} implementations  require consideration of \emph{how} clients will be able to inter-administratively learn keys by names.
In the case of RPKI, applying its namespace (\ie IP/ASN) to general objects of arbitrary types would be challenging.
However, DNSSEC is built on the Internet's de facto namespace (the DNS).
In fact, as outlined in our prior work~\cite{osterweil2020cybersecurity}, the DNS-based Authentication of Named Entities (DANE)~\cite{rfc6698,rfc7671} protocol suite is designed to generalize origin authentication to other protocols.
Examples include extending DANE protections for S/MIME~\cite{rfc8162} and PGP~\cite{rfc7929}.
With DNSSEC's general namespace and those protocols' application to arbitrary object types, the ingredients for generalizable object security are (arguablly) already operational!

Similarly, DNSSEC, RPKI, and Authenticode include lifetime protections.
Though, none of these include guidance or intuition on how to \emph{meaningfully} set these lifetimes.
DNSSEC and RPKI  are soft-state protocols, which described as protecting against replay attacks.
Authenticode implements this protection as optional, and off-by-default.

Interestingly, none of these protocols fully specifies the keys-to-objects protections, \emph{except} Authenticode.
In that protocol, the management of keys' lifecycles is (by default) decoupled from objects by using Time Stamp Authorities (TSAs).
The generalizability and robustness of this approach may be open to debate, but it is a rare example where this protection has been inter-administratively operationalized. 

Integrity protections have been implemented by all of the protocols we considered.
However, provenance has only been protected by C2PA.
The type of provenance that C2PA protects involves the modification history of media objects.
As we noted throughout, our view is that a more general form (process provenance) will be necessary, in order to  expose the derivation history and relationships of objects.
Further, as data objects now have their own \emph{agency}, additional provenance showing objects' chain of custody (\ie where have objects been circulated) will be very important for critical thinking frameworks.
The technical mechanism by which C2PA implements its provenance lends itself very well to both the protections it is designed for \emph{and} also potential extensibility to the type of process provenance we are suggesting.

Finally, the universal lack of evaluability to the object-level protections of these protocols is a critical absence.
This is, perhaps, also underscored by the expansive evaluations of the properties that are \emph{derived} on top of these protocols (as described in Section~\ref{sec:past}).
We argue that what is needed for a generalized object security service layer is an evaluation approach that can quantify \emph{the protections on the objects themselves}.

\para{\bf Ways Forward:}
Robust object security protections will face usability challenges.
The complex interplay between protections may face adoption challenges if, for users to benefit from them, they require inherit complexity.
For example, if users must manually inspect process provenance trees to determine if a news story is derived from sources they trust, the protections may become too cumbersome.
Moreover, if a single (secure) news story reports a view that differs from a corpus of non-secured news stories, how should users react?
For the former concern, we envision an important vein of usable security research, whereby complexity can be automated (while still being available for inspection).
For example, consider a process provenance graph which shows the sources of information in its digital objects, and details where it has been promoted and seen online (including all of the embedded images and source images).
Large Language Models (LLMs) have become a growing staple of Internet users who seek summaries of complex information.
Providing an object secured input would allow complex interpretation of both objective and, potentially, subjective aspects of critical thinking frameworks.
Furthermore, a cryptogtaphically verifiable credential from the specific LLM used would then need to be added to a digital object's process provenance.
This would still allow a user to visually inspect an object's security properties, but serve to streamline the \emph{usage}.
Additionally, consider training corpuses for models that online contain object secured data.
In the case of the latter concern, we propose that discovering a path towards a consensus view of reality from digital objects will best be found through approaches.

\para{\bf A Tactical Road Map:}
We propose that existing general-by-design protections lay a tactical path forward.
Whereas our Table~\ref{tab:synth} shows that no single object security protocol deployed today implements all of the security services described, several implement generalizable protections and can (and should) be combined.
By using S/MIME (or PGP), DNSSEC, and C2PA together, the combined protections would include general: origin authentication, integrity, provenance, and confidentiality.
Providing the linkage for this would seem to be the basic challenge, but this is precisely what the DANE protocol suite is designed for.
RFCs~\cite{rfc8162, rfc7929} specify the protocols for binding DNSSEC's origin authentication to S/MIME and PGP, respectively.
These protocols also serve as ready templates for extending origin authentication protections from DNSSEC to C2PA, thereby enhancing its provenance protections.
Lastly, C2PA's protocol bases its provenance protections on the general-purpose JUMBF, whose extensiblilty will allow enhanced process provenance protections.

%% file: conc.tex
\section{A Call to Arms}\label{sec:conc}

We are issuing a \emph{call to arms:} the era of unsecured digital objects on the Internet must be constrained to rapidly restore our ability to integrate \emph{genuine} online facts to support resilient cognition and robust decision-making. To turn the tide against this Cyber-Psychosis, a first step is to develop a foundational understanding of \emph{how} to effectively implement generalized object security in a way that supports rational critical thinking even with the prevalence of deceptive online content.
Only by establishing a foundation of secure objects can we then hope to develop a cognitive sense of consensus reality from online sources and overcome Cyber-Psychosis.

We propose that the Internet needs a \emph{generalized object security service layer}, to secure digital objects \emph{by default}.
As a straw man, in Section~\ref{sec:objs} we proposed six candidate types of protections needed: \one origin authentication, \two lifetime protections, \three key/object lifecycle management, \four integrity and provenance, \five confidentiality, and \six evaluability.
However, we \emph{already have} been deploying many of the basic pieces that we need (in some cases for decades). What stands as our challenge to the community is to \emph{use} what we have, and develop any remaining pieces. Origin authentication already exists in large-scale inter-administrative protocols like DNSSEC and RPKI. What's more, DNSSEC is a generalized namespace and multiple DANE protocols have bridged it to general objects, specifically using S/MIME and PGP, as previously observed~\cite{osterweil2020cybersecurity}. Yet, C2PA (which has produced a foundationally advanced provenance framework) has not yet embraced full origin authentication. We believe that protocols like C2PA can easily be evolved to use DANE, thereby enhancing them one step further toward \emph{full object security} protocols!

Furthermore, while lifetime protections have been developed in multiple protocols, we need to backstop them with guidance about \emph{how to use them effectively}. Our protocols that make use of public key cryptography have struggled with key-to-object lifecycle alignment. That said, Authenticode has produced one approach that manages this, and recent research in DNSSEC~\cite{osterweil2022beginning} illustrates an alternative methodology to managing this. Perhaps the final missing piece is \emph{evaluability} of object protections. However, we contend that this is \emph{not} a blocking precondition. The necessary basic research will be aided by pursuing holistic deployments and interdisciplinary efforts which tie an understanding into resilient critical thinking with object security and technical developments. Individually we have many of the necessary pieces, we need to reach across our disciplinary research silos and work \emph{together!}